\documentclass[twocolumn,showpacs,preprintnumbers,amsmath,amssymb,prb]{revtex4}
\usepackage[colorlinks, linkcolor=blue,anchorcolor=blue,citecolor=blue,urlcolor=blue]{hyperref}
\usepackage{graphicx}
\usepackage{dcolumn}
\usepackage{bm}
\usepackage{epstopdf}

\begin{document}

\title{
Dynamical Correlation Functions of the Quadratic Coupling Spin-Boson Model }

\author{Da-Chuan Zheng}
\affiliation{Department of Physics, Renmin University of China, 100872 Beijing, China}
\author{Ning-Hua Tong}
\email{nhtong@ruc.edu.cn}
\affiliation{Department of Physics, Renmin University of China, 100872 Beijing, China}
\date{\today}

\begin{abstract}
The spin-boson model with quadratic coupling is studied using the bosonic numerical renormalization group method. We focus on the dynamical auto-correlation functions $C_{O}(\omega)$, with the operator $\hat{O}$ taken as $\hat{\sigma}_x$, $\hat{\sigma}_z$, and $\hat{X}$, respectively. In the weak-coupling regime $\alpha < \alpha_c$, these functions show power law $\omega$-dependence in the small frequency limit, with the powers $1+2s$, $1+2s$, and $s$, respectively. At the critical point $\alpha = \alpha_c$ of the boson-unstable quantum phase transition, the critical exponents $y_{O}$ of these correlation functions are obtained as $y_{\sigma_x} = y_{\sigma_z} = 1-2s$ and $y_{X}=-s$, respectively. Here $s$ is the bath index and $X$ is the boson displacement operator. Close to the spin flip point, the high frequency peak of $C_{\sigma_{x}}(\omega)$ is broadened significantly and the line shape changes qualitatively, showing enhanced dephasing at the spin flip point.

\end{abstract}
\pacs{05.10.Cc, 64.70.Tg, 03.65.Yz, 05.30.Jp}

\keywords{ quadratic-coupling spin-boson mode, numerical renormalization group, quantum phase transition, dynamical correlation function}

\maketitle

\begin{section}{Introduction }

The spin-boson model (SBM) is a frequently used paradigm to study the influence of the environmental noise on the quantum evolution of a two-level system.\cite{Leggett1,Weiss1} The environment-induced dissipation and dephasing are the key issues in many fields of physics, ranging from biology to the endeavour of building a quantum computer.\cite{Caldeira1} Sufficiently strong  coupling to the boson bath also induces localize-delocalize quantum phase transitions (QPTs) in the two-level system for the Ohmic and sub-Ohmic baths.\cite{Leggett1,Weiss1,Kehrein1,Bulla1,Bulla1p} The non-trivial universality class of these QPTs receives much attention in recent years\cite{Vojta1,Vojta2,Vojta2p,Vojta3,Winter1,Alvermann1,Guo1,Hou1,Tong1} and the experimental detection of such QPTs has been proposed.\cite{Recati1,Tong2,Tong2p,Orth1,Porras1}

In the conventional spin-boson model, a spin which describes the two-level quantum system is coupled linearly to the displacement operator of a group of harmonic oscillators which are used to describe the environmental noise. Recently, In the experiments of superconducting quantum bit (qbit) system, the linear qubit-environment coupling can be tuned to zero to suppress the decoherence. Significant enhancement of the coherence time was observed in experiments at such an optimal working point.\cite{Vion1,Bertet1,Ithier1,Yoshihara1,Kakuyanagi1} Motivated by these experimental progress, much attention is paid to the spin-boson model with a quadratic coupling which is the leading term at the optimal working point. Theoretical studies of the quadratic-coupling spin-boson model (QSBM) have been carried out, mainly focusing on the dissipation and decohrence of the qubit.\cite{Makhlin1,Bergli1,Cywinski1,Balian1}. QSBM is also studied in other contexts such as the quantum dot-based qubit systems\cite{Muljarov1,Borri1} and the quantum Brownian motion of a heavy particle.\cite{Maghrebi1}

In a recent work,\cite{Zheng1} we studied the zero temperature properties of the sub-Ohmic QSBM using the numerical renormalzation group (NRG) method.\cite{Wilson1,Bulla2} We found that the bosonic environment is unstable towards local distortion under sufficiently strong quadratic spin-boson coupling, resulting in a novel impurity-induced environmental QPT in this model. We produced a ground state phases diagram on the $\epsilon$ (bias) -$\alpha$ (coupling strength) plane which contains both the continuous and the first-order QPTs. On this phase diagram, $\langle \sigma_z\rangle$ changes sign at the so-called spin-flip line which is a first-order transition line at $\Delta=0$ but becomes a continuous crossover line at $\Delta > 0$ ($\Delta$ is the tunnelling strength of the quantum two-level systems, see below.). The critical exponents of the QPT are obtained exactly from the exact solution at $\Delta=0$. The equilibrium dynamical correlation functions of the z-component of spin $C_{\sigma_{z}}(\omega)$, and that of the bath displacement operator $C_{X}(\omega)$ are analysed near the QPT. These results disclosed the strong impurity-bath mutual influence due to the non-linearity of the coupling.

In this paper, using NRG, we explore the evolution of the equilibrium dynamical correlation functions $C_{\sigma_{x}}(\omega)$ as the parameters $\epsilon$ and $\alpha$ are tuned throughout the phase diagram. To make comparisons, we also summarize the results for $C_{\sigma_{z}}(\omega)$ and $C_{X}(\omega)$ which have been obtained in Ref.~\onlinecite{Zheng1}. These correlation functions reflect important properties of the model. $C_{\sigma_{x}}(\omega)$ and $C_{\sigma_{z}}(\omega)$ contains information about dissipation and decoherence of the spin, respectively. $C_{X}(\omega)$ characterized the bath which is severely influenced by the presence of the impurity in the case of the non-linear coupling.
We find that in the weak-coupling regime $\alpha < \alpha_c$, all the three correlation functions have power law form in the small frequency limit. $C_{X}(\omega) \sim \omega^{s}$ reflects the power law spectral function of the bath that we used $J(\omega) \propto \alpha \omega^{s}$.  $C_{\sigma_x}(\omega)$ is similar to $C_{\sigma_z}(\omega)$\cite{Zheng1}, with the form $C_{\sigma_{x}}(\omega) \sim \omega^{1+2s}$.  At the critical point $\alpha = \alpha_c$, $C_{O}(\omega) \propto \omega^{y_{O}}$ in the small frequency limit ($\hat{O} = \sigma_x$, $\sigma_z$, and $X$). We find  that $y_{\sigma_x}= y_{\sigma_z} = 1-2s$ and $y_{X} = -s$. In the higher frequency regime close to the Rabi frequency $\omega_R$, both $C_{\sigma_{x}}(\omega)$ and $C_{\sigma_{z}}(\omega)$ have a prominent peak even at the strongest coupling $\alpha = \alpha_c$ before the environment gets unstable. Close to the spin-flip line, a significant broadening of the Rabi peak is observed in $C_{\sigma_{x}}(\omega)$ but not in $C_{\sigma_{z}}(\omega)$, showing an enhanced decoherence at the spin-flip line of the optimal working point.

This paper is organized as the following. In section 2 we introduce QSBM and the formalism we used to calculate the equilibrium correlation function with bosonic NRG method. Section 3 presents results from our NRG study. A conclusion is given in section 4.

\end{section}

\begin{figure}
 \vspace*{-0.0cm}
\begin{center}
  \includegraphics[width=240pt, height=180pt, angle=0]{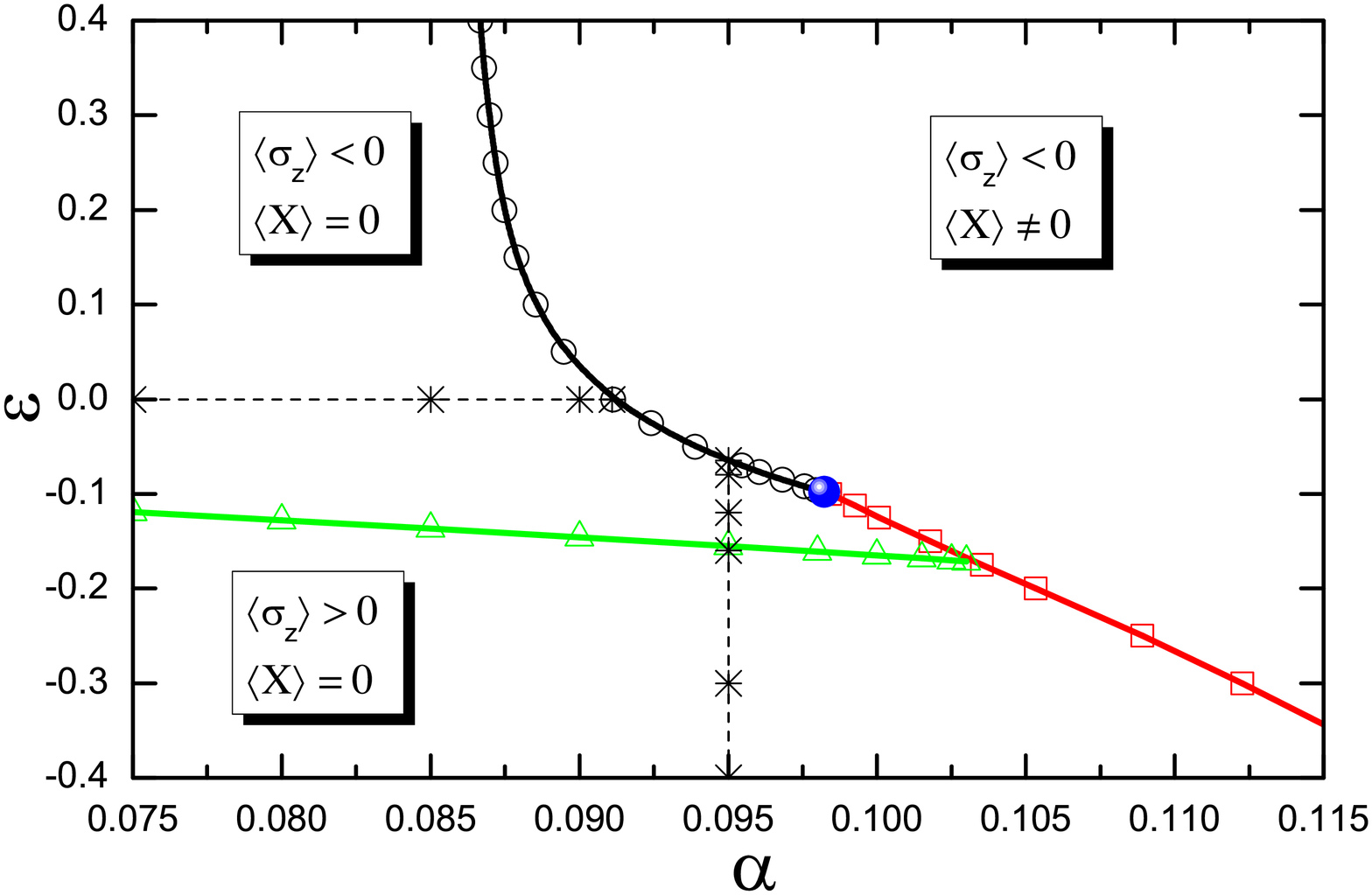}
  \vspace*{-0.5cm}
\end{center}
  \caption{(color online) Phase diagram of QSBM for $s=0.3$ and $\Delta=0.1$ obtained from NRG. The circles, squares, and up triangles with guiding lines represent continuous QPT, first-order QPT, and the spin flip lines, respectively. The solid dot marks the jointing point of the continuous and the first-order QPTs. The stars with dashed lines are the points for which the correlation functions are calculated in Fig.2 (horizontal line) and Fig.3 (vertical line). NRG parameters are $\Lambda=4.0$, $M_s=100$, and $N_{b}=12$.
}   \label{Fig1}
\end{figure}

\begin{section}{Model and Method }

The Hamiltonian of the quadratic coupling spin-boson model reads
\begin{equation}
  H_{QSB} = \frac{\epsilon}{2} \sigma_z -\frac{\Delta}{2} \sigma_x + \displaystyle\sum_{i} \omega_{i} a_{i}^{\dagger}a_{i} + \frac{g_2}{2}  \sigma_{z} \hat{Y}^{2}.
\end{equation}
The two-level system is described by a spin-1/2 operator with the bias $\epsilon$ and tunnelling strength $\Delta$. It is coupled to the  bosonic bath with mode energies $\{ \omega_i \}$ via the local boson displacement operator $\hat{Y} = \sum_{i} \lambda_{i}(a_{i}+a_{i}^{\dagger})$. $g_2$ is the second-order coefficients in the expansion of a general coupling form $\sigma_z f(\hat{Y}) = \sigma_z (g_0 + g_1\hat{Y} + g_2 \hat{Y}^{2} + ...)$. At the optimal working point of the superconducting qubit circuit, $g_1=0$ and the remaining leading order coupling is $g_2$. The effect of the bath on the spin is encoded into the bath spectral function
\begin{equation}
   J(\omega) \equiv \pi \sum_{i} \lambda_{i}^{2} \delta(\omega -   \omega_{i}).
\end{equation}
Although the QPT exists also for the single mode quadratic coupling Hamiltonian which is relevant to the qubit-resonator system,\cite{Niemczyk1,Diaz1} in this paper we mainly focus on the continuous bath with a power law spectrum in the small $\omega$ limit and a hard-cutoff at $\omega = \omega_c$,
\begin{equation}
   J(\omega) = 2\pi\alpha \omega^{s} \omega_{c}^{1-s}, \,\,\,\,\, \,\,\,\, (0 \leqslant \omega  \leqslant   \omega_c).
\end{equation}
Here $s \geqslant 0$ is the exponent of the bath spectrum and $\alpha$ controls the strength of the spin-boson coupling. We set $\omega_c = 1.0$ as the unit of energy. The quadratic coefficient $g_2$ can be absorbed into $\lambda_i$'s and for convenience we set it as unity. In this paper, we fix $\Delta=0.1$ to study the dependence of the equilibrium dynamics on $\epsilon$ and $\alpha$. We confine our study to the sub-Ohmic bath with $0 < s < 1$. For definiteness, we use a generic value $s=0.3$ in most part of this paper and discuss the extension of our conclusion to the full sub-Ohmic regime in the end.

We will use the bosonic NRG method to study this model. NRG is regarded as one of the most powerful methods for studying quantum impurity models because it is non-perturbative and reliable in the full parameter space.\cite{Wilson1,Bulla2} In general, the errors in the NRG calculation come from two sources. One is the approximation of using one bath mode to represent each energy shell, which is controlled by the logarithmic discretization parameter $\Lambda \geqslant 1$. The other is the truncation of the energy spectrum after each diagonalization to overcome the exponential increase of the Hilbert space, which is controlled by the number of kept states $M_s$. For the bosonic NRG, an additional source of error is the truncation of infinite dimensional Hilbert space of each boson mode into $N_b$ states on the occupation basis. Exact results can be obtained only in the limit $\Lambda=1$, $M_s= \infty$, and $N_b = \infty$. Details of the application of NRG to $H_{QSB}$ can be found in Ref.~\onlinecite{Zheng1}.

In Fig.1, we reprint the NRG phase diagram for $s=0.3$ from Ref.~\onlinecite{Zheng1}. It is obtained using the NRG parameters $\Lambda=4.0$, $M_s=100$, and $N_b=12$. Though the phase boundaries depend quantitatively on the NRG parameters, qualitative topology of the phase diagram is unchanged when we extrapolate $\Lambda$, $M_s$, and $N_b$ to the exact limit.\cite{Zheng1} In Fig.1, the ground state of $H_{QSB}$ in the $\epsilon$-$\alpha$ plane is characterized by two physical quantities, $\langle \sigma_z\rangle$ and $\langle \hat{X} \rangle$. Here $\hat{X} \equiv \hat{Y}/ \sqrt{ \sum_{i}\lambda_i^{2} }$ is the normalized local boson displacement operator. On the left-hand side of the phase diagram (smaller $\alpha$), the symmetry of $H_{QSB}$ allows $\hat{X}$ to fluctuate symmetrically around zero while keeping the spin polarized, giving $\langle X \rangle =0$. On the right-hand side of the phase diagram (larger $\alpha$), the coupling is sufficiently strong so that the harmonic oscillators in the environment get softened and inversion of the harmonic potentials of the low energy modes occurs, leading to the breaking of the symmetry of $H_{SQB}$ and $\langle X \rangle \neq 0$. Therefore, $\langle X \rangle$ is the order parameter of this environmental QPT. The QPT between the two phases is continuous (empty circles with eye-guiding line) for larger $\epsilon$ and is of first order (empty squares with eye-guiding line) for smaller $\epsilon$. At the first-order QPT, $\langle X \rangle$ jumps abruptly from zero to a finite value. We denote the critical coupling strength as $\alpha_c^{(c)}$ for the former and $\alpha_{c}^{(1)}$ for the latter. These two kinds of QPT lines meet at the jointing point ($\alpha_c$, $\epsilon_c$) (solid circle in Fig.1) where the finite jump in $\langle X \rangle$ at $\alpha_{c}^{(1)}$ shrinks to zero. For the Hamiltonian Eq.(1), $\langle X \rangle = \pm \infty$ in the environment-unstable phase. If the higher order anharmonic terms of the bosons beyond Eq.(1) is considered, $\langle X \rangle$ will be confined to a finite value in the boson-unstable phase. The boson-unstable phase then describes the physical situation where the environmental degrees of freedom have a local distortion around the impurity.

Since the instability of bosons in the strong coupling regime only occurs when the pre-factor of $\hat{Y}^{2}$ is negative in the coupling term of $H_{QSB}$, the above QPTs occur either with $\langle \sigma_{z} \rangle < 0$ on both sides of the transition (upper part of Fig.1), or with $\langle \sigma_{z} \rangle$ changing from positive to negative (lower part of Fig.1). In the boson-stable regime, there is a spin flip line $\epsilon_f(\alpha)$ (up triangles with eye-guiding line) which separates $\langle \sigma_{z} \rangle < 0$ from $\langle \sigma_{z} \rangle > 0$. For $\Delta \neq0$, $\langle \sigma_z \rangle$ continuously crosses zero at this line.

At $\Delta=0$, $H_{QSB}$ is exactly soluble because the two subspaces with $\sigma_z = \pm 1$ are decoupled.\cite{Zheng1} The phase diagram is qualitatively the same as the one for $\Delta >0$ shown in Fig.1. The exact critical coupling strength is obtained as $\alpha_{c}^{(c)} = \alpha_c^{(1)} = s/(4g_2 \omega_c)$. The spin flip line is the exact level crossing line between the two subspaces with $\sigma_z = \pm 1$. To leading order of $\alpha$, the spin flip line is given by $\epsilon_f = -2 \alpha / (1+s) (g_2 \omega_c^{2})$. For a finite $\Delta$, NRG study showed that the spin flip line becomes a smooth crossover line without singularity.

Below, in order to further explore the dissipation and decoherence of the quantum two-level system subjected to the influence of a bath with quadratic coupling, we study the dynamical correlation function at $T=0$ using NRG. The dynamical correlation function of an operator $\hat{O}$ is defined as
\begin{equation}
   C_{O}(t) \equiv \frac{1}{2} \langle [ \hat{O}(t), \hat{O}(0) ]_{+}\rangle ,
\end{equation}
where $[\hat{A}, \hat{B} ]_{+} = \hat{A}\hat{B} + \hat{B}\hat{A}$ is the anti-commutator of $\hat{A}$ and $\hat{B}$. What we calculate directly is its Fourier transformation
\begin{equation}
   C_{O}(\omega) \equiv \frac{1}{2\pi} \int_{-\infty}^{+\infty} C_{O}(t) \,dt.
\end{equation}
In this paper, we study $C_{O}(\omega)$ at zero temperature for the following operators $\hat{O} = \sigma_x$, $\hat{O} = \sigma_z$, and $\hat{O} = X$. For a non-degenerate ground state, the Lehmann representation of $C_{O}(\omega)$ at $T=0$ is written as
\begin{eqnarray}
   C_{O}(\omega) &=& \frac{1}{2} \sum_{n} | \langle 0| \hat{O} |n \rangle |^2 \delta(\omega+ E_{0}-E_{n})   \nonumber \\
   && +  \frac{1}{2} \sum_{n} | \langle n| \hat{O} |0 \rangle |^2 \delta(\omega+ E_{n}- E_{0}).
\end{eqnarray}
Here $|n \rangle$ and $E_n$ are the $n$-th eigen state and energy of the Hamiltonian, respectively. In general, $C_{O}(\omega)= A \delta(\omega) + C^{\prime}_{O}(\omega)$, where $A= |\langle 0| \hat{O} |0 \rangle|^{2}$ and $|0 \rangle$ is the ground state.
$C_{O}(\omega)$ is an even function of $\omega$ and it fulfils the sum rule
\begin{equation}
   \int_{-\infty}^{+\infty} C_{O}(\omega)  \, d \omega =  \langle 0 | \hat{O}^{2} | 0 \rangle.
\end{equation}

Within NRG, $C_{O}(\omega)$ can be calculated using the patching method.\cite{Bulla3} In this method, the poles and weights obtained from the diagonalization of each Wilson chain Hamiltonian $H_{N}$ are collected and patched together, to form a full spectrum ranging from high energy to the lowest energy reachable by NRG. The obtained $\delta$-peaks are then broadened using the log-Gaussian function with a broadening parameter $B$. In the patching method, the sum rule of $C_{O}(\omega)$ is fulfilled approximately, with a relative error at the level of a few percent. The more sophisticated full density matrix method\cite{Peters1,Weichselbaum1} can conserve the sum rule exactly but the positivity of $C_{O}(\omega)$ is not guaranteed.
In general, NRG method can give rather accurate low frequency spectral function, but the high frequency part is less reliable due to the loss of energy resolution from logarithmic discretization and the over broadening of the log-Gaussian function. There are attempts to improve the high frequency resolution within NRG method\cite{Oliveira1,Oliveira2,Bulla4,Campo1,Zitko1,Zitko2,Florens1,Osolin1} with various extent of success. In this paper, we calculate the above-stated correlation functions using the patching method with a broadening parameter $B=1.0$.
To improve the resolution of certain high frequency features, we use the $z$-average method\cite{Oliveira1,Oliveira2} with the number of $z$ values $N_z=10$ and a reduced broadening parameter $B=0.3$.

\end{section}

\begin{figure}
\vspace*{-0.0cm}
\begin{center}
  \includegraphics[width=240pt, height=240pt, angle=0]{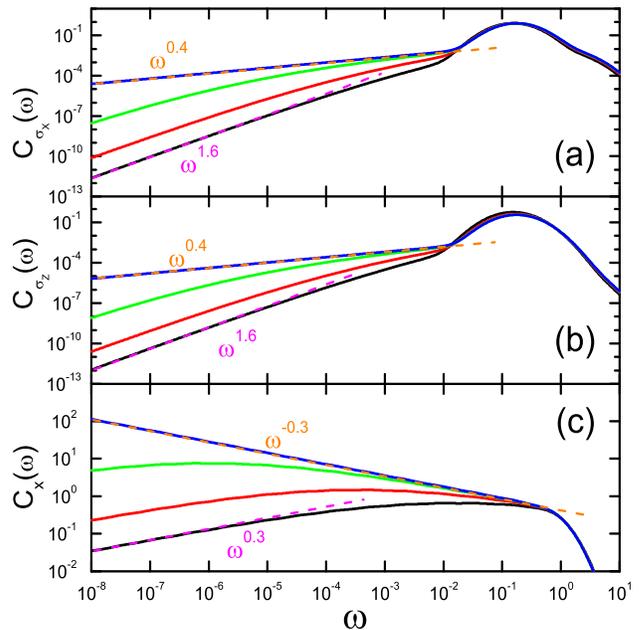}
  \vspace*{-0.5cm}
\end{center}
  \caption{(color online) Equilibrium auto-correlation functions of the operators $\sigma_{x}$ (a), $\sigma_z$ (b), and $\hat{X}$ (c), obtained from NRG for $s=0.3$, $\Delta=0.1$, and $\epsilon=0.0$. In each figure, from bottom to top, $\alpha=0.075$, $0.085$, $0.09$, and $0.0911 \approx \alpha_{c}^{(c)}$, corresponding to the stars along the horizontal dashed line in Fig.1. The dashed lines are power functions of $\omega$ with the prescribed exponents. For (b), the zero frequency peak $A \delta(\omega)$ is not shown. NRG parameters are same as Fig.1 and the broadening parameter $B=1.0$.
}  \label{Fig2}
\end{figure}

\begin{figure}
  \vspace*{-0.0cm}
\begin{center}
  \includegraphics[width=240pt, height=240pt, angle=0]{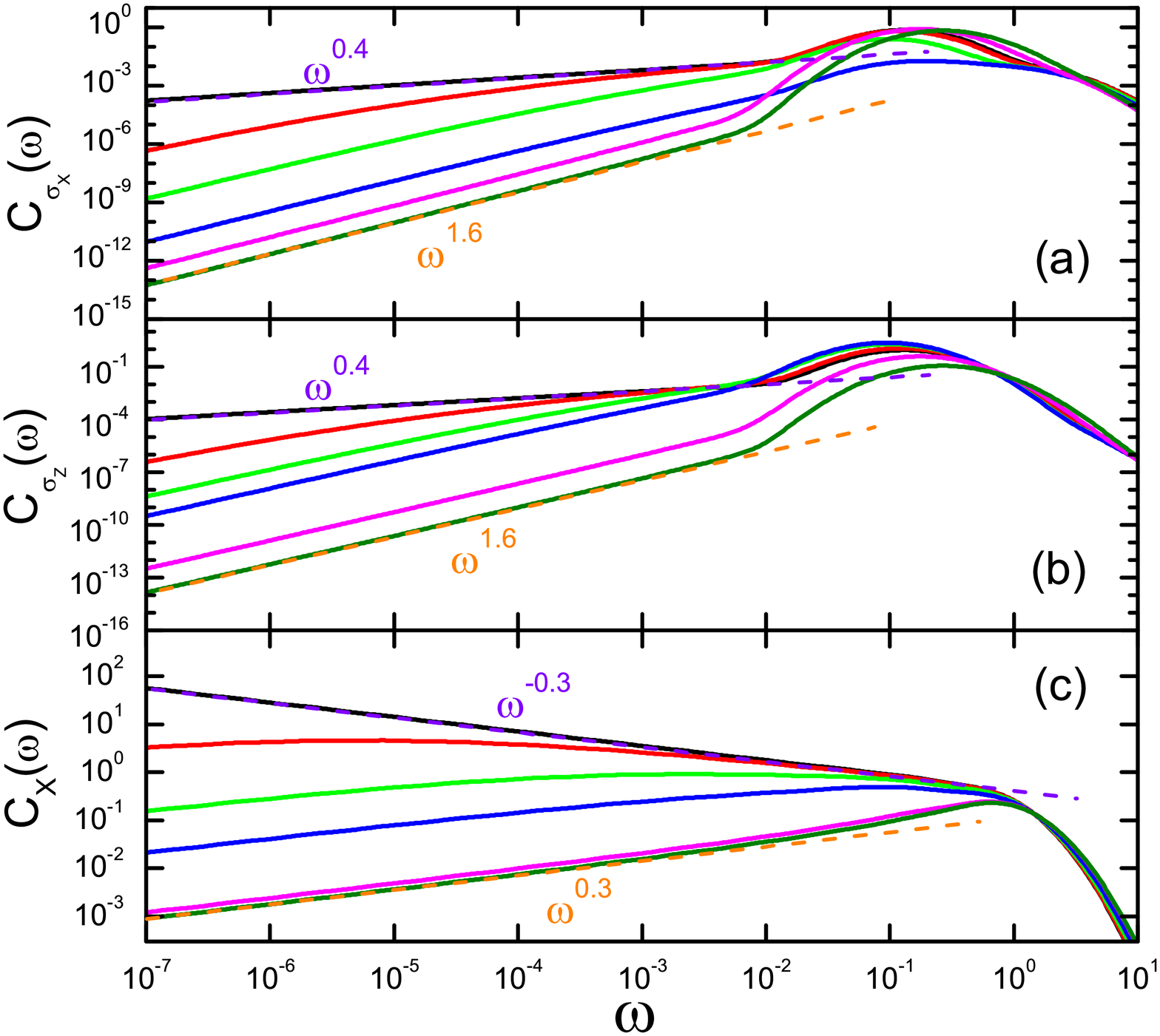}
  \vspace*{-0.5cm}
\end{center}
  \caption{(color online) Equilibrium auto-correlation functions of the operators $\sigma_{x}$ (a), $\sigma_z$ (b), and $\hat{X}$ (c), obtained from NRG for $s=0.3$, $\Delta=0.1$, and $\alpha=0.095$. In each figure, from bottom to top in the small frequency regime, $\epsilon=-0.4$, $-0.3$, $-0.16$, $-0.12$, $-0.08$, and $-0.06482 \approx \epsilon_{c}^{(c)}$, corresponding to the stars along the vertical dashed line in Fig.1. The dashed lines are power functions of $\omega$ with the prescribed exponents. For (b), the zero frequency peak $A \delta(\omega)$ is not shown. NRG parameters are same as Fig.1 and the broadening parameter $B=1.0$.
}    \label{Fig3}
\end{figure}

\begin{figure}
\vspace*{-0.0cm}
\begin{center}
  \includegraphics[width=240pt, height=200pt, angle=0]{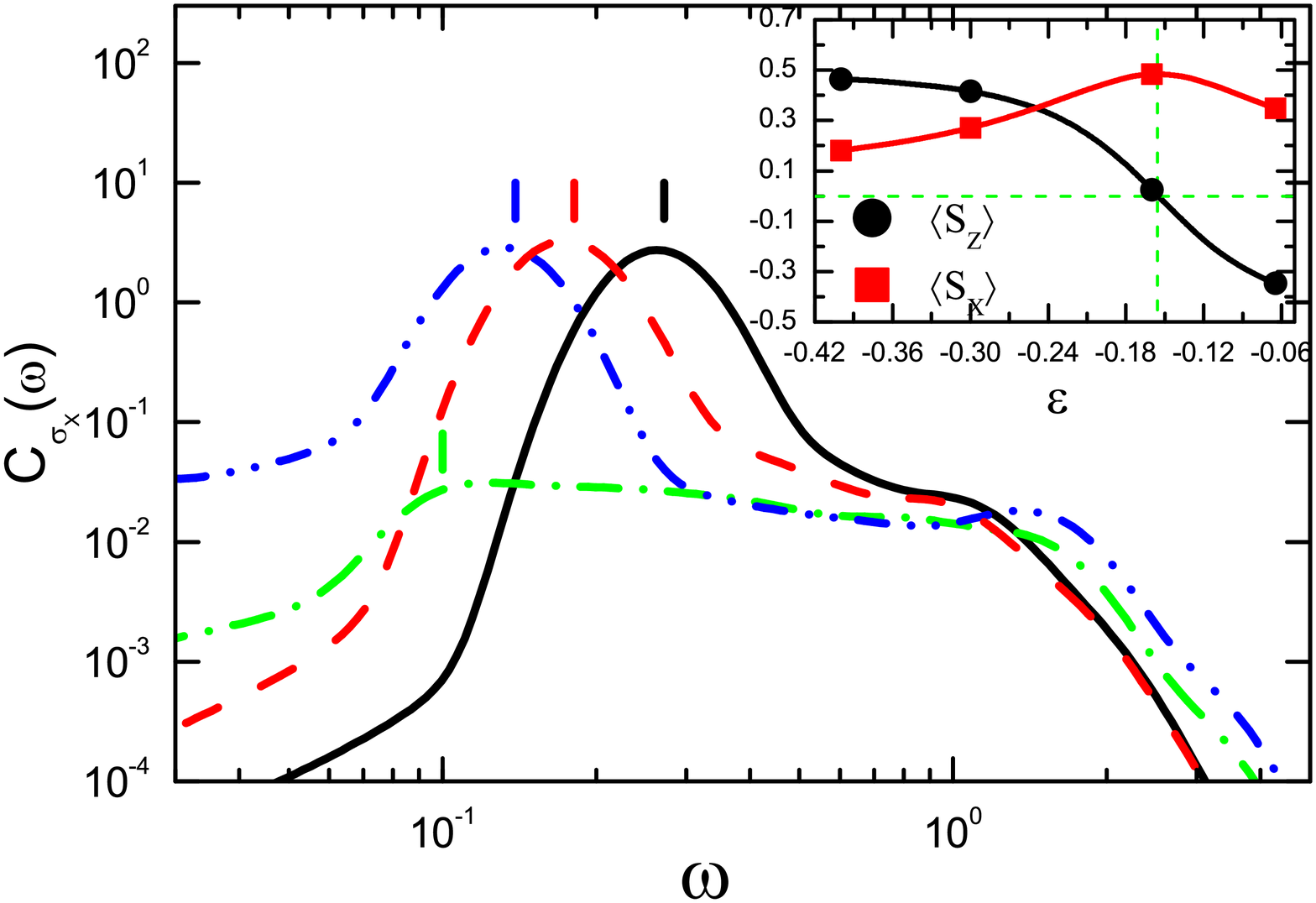}
  \vspace*{-1.0cm}
\end{center}
  \caption{(color online)
  High frequency peak of $C_{\sigma_{x}}(\omega)$ for $s=0.3$, $\Delta=0.1$, $\alpha=0.095$, and various $\epsilon$ values. The solid, dashed, dash-dot, and dash-dot-dot lines are for $\epsilon= -0.4$, $-0.3$, $-0.16$, and $-0.06482$, respectively. The vertical dashes mark out the fitted Rabi frequency $\omega_{R}$ (see text). Inset: $\langle S_z \rangle$ (circles) and $\langle S_x \rangle$ (squares) as functions of $\epsilon$. The symbols are results for the corresponding $\epsilon$ values of the main figure. The dashed lines mark out the spin flip point where $\langle S_{z}\rangle $ changes sign. NRG parameters are same as Fig.1. The broadening parameter $B=0.3$ and $C_{\sigma_{x}}(\omega)$'s are averaged over $N_z=10$ different $z$ values.
}     \label{Fig4}
\end{figure}

\begin{section}{Results}

In order to explore the evolution of the dynamical correlation functions $C_{O}(\omega)$ (for $\hat{O}= \sigma_x$, $\sigma_z$, and $X$) with $\epsilon$ and $\alpha < \alpha_c$, we scan the parameters along two representative paths on the $\epsilon$-$\alpha$ plane. First, we fix $\epsilon=0.0$ and increase $\alpha$ (stars along the horizontal dashed line in Fig.1). Second, we fix $\alpha =0.095$ and increase $\epsilon$ (stars along the vertical dashed line in Fig.1). Both paths are confined in the boson-stable phase. The obtained $C_{O}(\omega)$'s are plotted in Fig.2 and Fig.3, respectively, for making systematic comparisons.

In Fig.2(a)-(c), we show $C_{O}(\omega)$ on the logarithmic scale calculated at $\epsilon=0$ and for a series of $\alpha$ values. $\hat{O} = \sigma_x$, $\sigma_z$, and $X$ for Fig.2(a), (b), and (c), respectively. In each figure, the curves from bottom to top correspond to increasing $\alpha$ values. For such a series of parameters, $C_{\sigma_z}(\omega)$ and $C_{\sigma_x}(\omega)$ are quantitatively similar, both composed of the low frequency power law behavior and the high frequency peak around $\omega \sim 0.1$. The broad peak is to much extent due to the log-Gaussian broadening of an actually much narrower Rabi coherent peak. The fitted values of the low frequency exponents are numerically close for $C_{\sigma_z}(\omega)$ and $C_{\sigma_x}(\omega)$, being approximately $1.6$ for $\alpha < \alpha_c$ and $0.4$ for $\alpha = \alpha_c$. $C_{X}(\omega)$ has a similar behavior but lacks the coherent peak around the Rabi frequency. The corresponding exponents are close to $0.3$ and $-0.3$, respectively. The calculation for other $s$ values shows that the exponents of $C_{\sigma_z}(\omega)$ and $C_{\sigma_x}(\omega)$ agree with the expressions $1+2s$ for $\alpha < \alpha_c$ and $1-2s$ for $\alpha = \alpha_c$. The corresponding exponents of $C_{X}(\omega)$ are $s$ and $-s$, respectively, as being obtained exactly in Ref.~\onlinecite{Zheng1}.

We note that as $\alpha$ approaches $\alpha_c$ from below, the static averages $\langle \sigma_x \rangle$ and $\langle \sigma_z\rangle$ changes very slowly and they have negligible influence on the evolution of correlation functions. In this process, the crossover scale $\omega^{\ast}$ separating the high frequency $\omega^{1-2s}$ to the low frequency $\omega^{1+2s}$ behavior decreases to zero as $\omega^{\ast} \propto (\alpha_c - \alpha)^{z\nu}$, with $z=1$ being the dynamical exponent and $\nu$ the correlation length exponent (Ref.~\onlinecite{Zheng1}). In contrast, The high frequency peaks of both $C_{\sigma_z}(\omega)$ and $C_{\sigma_x}(\omega)$ do not change much. This stability of the peak position and line shape of the coherent peak implies a robust coherent short-time quantum evolution and the weak dissipation and docoherence effects near the quantum critical point.

\begin{figure}
\vspace*{-0.0cm}
\begin{center}
  \includegraphics[width=240pt, height=270pt, angle=0]{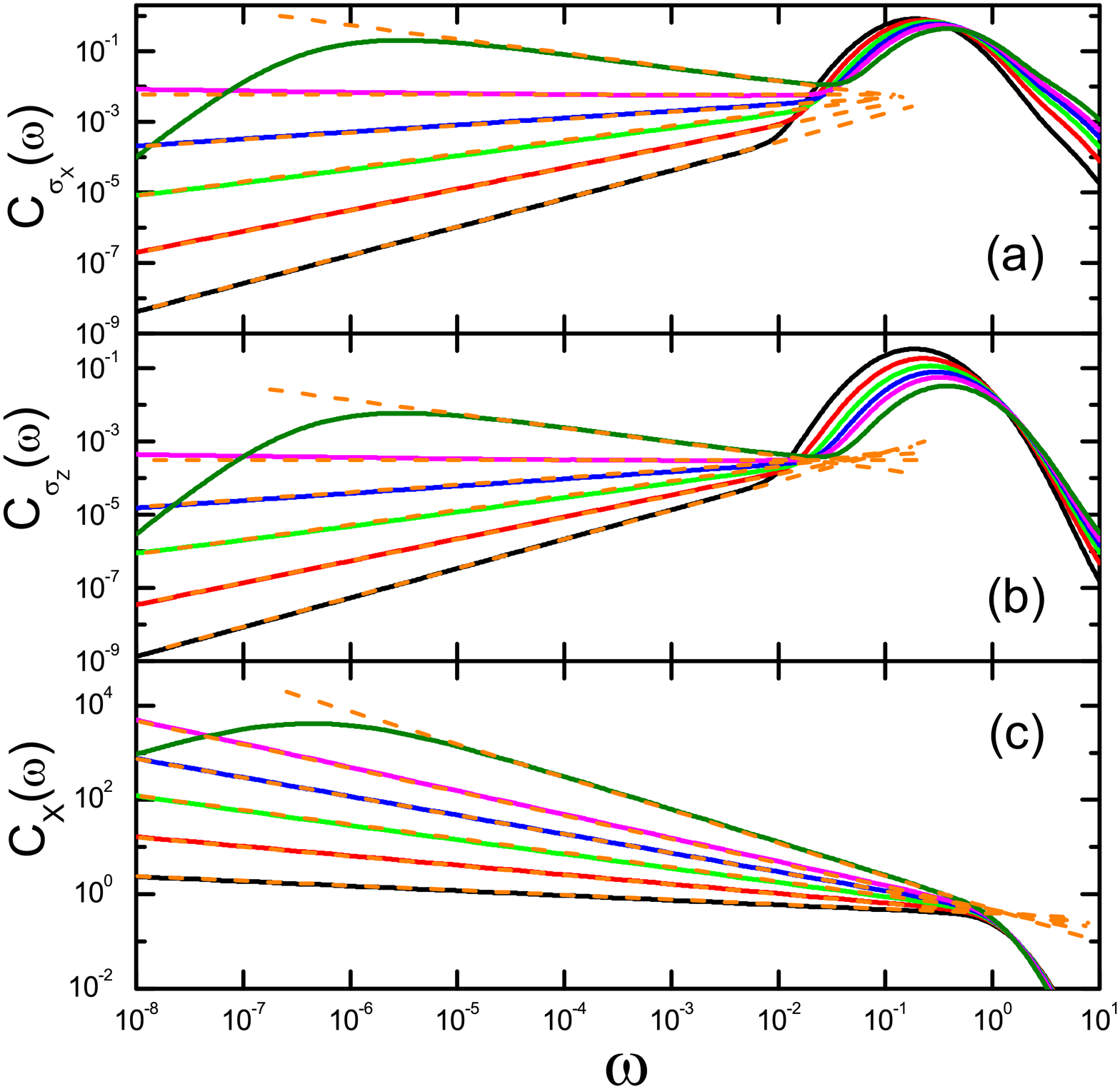}
  \vspace*{-0.5cm}
\end{center}
  \caption{(color online) Equilibrium auto-correlation functions of the operators $\sigma_{x}$ (a), $\sigma_z$ (b), and $\hat{X}$ (c), obtained from NRG for a series of $s$ values, $\Delta=0.1$, $\epsilon=0.1$, and $\alpha=\alpha_c(s)$. For each figure, from bottom to top in the small frequency regime, $s=0.1$, $0.2$, $0.3$, $0.4$, $0.5$, and $0.7$. The dashed lines are power functions of $\omega$ with the exponents $1-2s$ in (a) and (b), and $-s$ in (c). For (b), the zero frequency peak $A \delta(\omega)$ is not shown. NRG parameters are same as Fig.1 and the broadening parameter $B=1.0$.
}    \label{Fig5}
\end{figure}

In Fig.3, to investigate the evolution of the correlation functions in a different path, we fix $\alpha = 0.095 < \alpha_c$ and increase $\epsilon$ up to the QPT line. Compared to the evolution shown in Fig.2, some interesting features are observed here. First, the low frequency behavior is similar to that of Fig.2, {\it i.e.}, $C_{\sigma_x}(\omega)$ and $C_{\sigma_z}(\omega)$ have power law form $\omega^{1.6}$ for $\alpha < \alpha_c$ and the critical behavior $\omega^{0.4}$ at $\alpha = \alpha_c$. With increasing $\epsilon$, the low frequency value of $C(\omega)$ increases monotonically. $C_{X}(\omega) \propto \omega^{0.3}$ for $\alpha < \alpha_c$ and $C_{X}(\omega) \propto \omega^{-0.3}$ at $\alpha = \alpha_c$, with the crossover bahavior similar to that of Fig.2(c). The high frequency features are quite different from those of Fig.2. Here, the height and line shape of the high frequency peaks of $C_{\sigma_x}(\omega)$ and $C_{\sigma_z}(\omega)$ change significantly with increasing $\epsilon$, in contrast to those of Fig.2. Note that both Fig.3(a) and Fig.3(b) have $\delta$-peak at zero frequency, whose weights changes un-monotonically with $\epsilon$, as to be shown in Fig.4. As $\epsilon$ increases, the redistribution of weights between the zero-frequency $\delta$-peak and the finite frequency regime do account for part of the evolution shown in Fig.3(b). As $\epsilon$ increases from $-0.4$ to $-0.16$, $\langle S_z\rangle$ decreases from $0.5$ to almost zero (see inset of Fig.4). This leads to a decrease of the zero frequency weights and explains the uniform increase of $C_{\sigma_z}(\omega)$ in this $\epsilon$ regime, as shown in Fig.3(b). However, the line shape change of the high frequency peak in $C_{\sigma_x}(\omega)$ in Fig.3(a) cannot be simply attributed to it. This shows that the short-time decoherence in QSBM has some anomalous dependence on $\epsilon$ for a fixed coupling strength $\alpha$.

To further investigate the evolution of high frequency peak of $C_{\sigma_x}(\omega)$ with $\epsilon$, which is relevant to the dephasing time in the qbit experiment at the optimal working point, we use the $z$-average method to improve the frequency resolution. In Fig.4, we show the high frequency peak of $C_{\sigma_x}(\omega)$ at the same parameters as in Fig.3, for various $\epsilon$ values. Each curve is the averaged result over $N_z=10$ uniformly distributed $z$ values with a smaller broadening parameter $B=0.3$. Though the $z$-average method cannot remove all the errors of logarithmic discretization, it is argued to be useful for improving the resolution of high frequency features and remove the oscillations induced by a smaller broadening parameter.\cite{Oliveira1,Oliveira2} In the inset of Fig.4, the averages $\langle S_z \rangle$ and $\langle S_x \rangle$ are shown as functions of $\epsilon$ in the regime $-0.42 < \epsilon < -0.06428 \approx \epsilon_c$. The vertical dashed line marks out the spin flip point $\epsilon = \epsilon_f \approx -0.1556$. The horizontal dashed line marks out $\langle S_z \rangle=0$. Besides the shifts of peak position with $\epsilon$, significant change of the line shape of the high frequency peak in $C_{\sigma_x}(\omega)$ occurs close to the spin flip point $\epsilon_f \approx -0.16$. Close to this value of $\epsilon$, the high frequency peak is significantly broadened and the line shape is no longer a round peak. Away from this value of $\epsilon$, a relatively sharp peak appears at the effective Rabi frequencies $\omega_{R}$ on top of a broad background spectrum extending to $\omega \sim 10$.

The peak position is at the effective Rabi frequency  $\omega_R = \sqrt{\tilde{\epsilon} + \Delta_{r}}$. It can be estimated by assuming a free spin Hamiltonian $H_{eff} = -(1/2) \Delta_r \sigma_x + (1/2) \tilde{\epsilon} \sigma_z$. The effective bias $\tilde{\epsilon}$ contains both the original $\epsilon$ and the contribution from the static mean field of the quadratic coupling $(g_2/2) \sigma_z Y^{2}$ in $H_{QSB}$. $\Delta_r \approx \Delta$ is the renormalized tunnelling strength. We use the NRG data of $\langle S_z \rangle$ to solve for $\tilde{\epsilon}$ and then $\omega_R$. The obtained $\omega_R$'s (vertical dashes in Fig.4) agree well with the peak position in $C_{\sigma_x}(\omega)$, except for $\epsilon=-0.16$ close to the spin flip point. For $\epsilon=-0.16$, the fitted $\omega_R$ is located at the lower edge of a broad feature (which cannot be called a peak). It has been confirmed that at least for the weak coupling regime, the evolution of dynamical correlation function $C(t)$ is similar to $P(t)$, the average of the corresponding operator in the non-equilibrium situation \cite{Lv1}. A sharp high frequency peak in $C_{\sigma_x}(\omega)$ corresponds to coherent oscillations in $P_{\sigma_x}(t) = \langle \sigma_x(t)\rangle$ and it implies a weak dephasing effect. The vanishing of the peak close to the spin flip point at $\epsilon=-0.16$ thus corresponds to incoherent evolution with stronger dephasing effect.

In NRG, the $\delta$-peaks in the spectral function are obtained from iterative diagonalization of the logarithmically-discretized energy shells. They are then broadened by log-Gaussian functions. It is therefore difficult to tell the exact line shape of a high frequency peak from the raw NRG data, despite great efforts paid to improve on this problem.\cite{Oliveira1,Oliveira2,Bulla4,Campo1,Zitko1,Zitko2,Florens1,Osolin1} Here, however, the qualitative tendency gives a robust conclusion that close to the spin flip line, the decoherence peak in $C_{\sigma_x}(\omega)$ is broadened and the line shape changed qualitatively. This result, when translated into the non-equilibrium quantity, suggests that the dephasing time of the qbit may decrease significantly close to the spin flip line, being unfavorable to the realization of a long-lived qubit. Physically, from the point of vew of the weak $\Delta$ limit, the spin flip line at which $\langle \sigma_z \rangle =0$ can be approximately regarded as the degeneracy point of two (approximate) subspaces $\sigma_z = \pm 1$. The real quantum state is thus a superposition of $|\uparrow \rangle |\Psi_{\uparrow} \rangle$ and $| \downarrow \rangle|\Psi_{\downarrow} \rangle$ (two eigen states of $\sigma_z$) with equal weights. The energy fluctuations of these states due to coupling to bosons have a stronger effect in destroying the phase coherence in the superposition, leading to an enhanced dephasing at the spin flip point.

The above results are obtained for the sub-Ohmic bath with $s=0.3$. We have carried out systematic calculations for other sub-Ohmic $s$ values as well. In Fig.5, we show the dynamical correlation functions for a series of $s$ values at  $\alpha = \alpha_c(s)$, in order to find out the critical exponents $y_{O}$ defined as $C_{O}(\omega)(\alpha = \alpha_c) \propto \omega^{y_{O}}$. The fitted exponents from NRG data supports that $y_{\sigma_x} = y_{\sigma_z}$. Our results here confirm that $y_{\sigma_z} = 1-2s$ and $y_{X} = -s$ found in Ref.~\onlinecite{Zheng1}. For $\alpha < \alpha_c$, the power law in the low frequency limit are obtained as $C_{\sigma_x}(\omega) \propto \omega^{1+2s}$,  $C_{\sigma_z}(\omega) \propto \omega^{1+2s}$, and  $C_{X}(\omega) \propto \omega^{s}$ (data not shown). Close to the spin flip line, the significant broadening of the high frequency peak in $C_{\sigma_{x}}(\omega)$ and the change of the line shape are found to be universal for all sub-Ohmic regime.

Note that in Fig.5, the critical power law behavior for $s=0.7$ only appears in the high frequency regime $\omega \gg 10^{-6}$ and does not extend to arbitrarily small $\omega$. This is because for $s=0.7$, $\Delta=0.1$, and $\epsilon=0.1$, a first-order QPT occurs at $\alpha = \alpha_{c}^{(1)}$. It was found\cite{Zheng1} that as $s$ increases, the impurity-induced environmental QPT tends to become first order. Close to this QPT $\alpha = \alpha_c^{(1)}-0^{+}$, one can still observe the quantum criticality in the high frequency regime. This fact is interesting in that it shows an example of observing the quantum critical properties close to a weak first-order QPT which is much more prevailing than a continuous QPT in the experiments. Details of this issue will be studied elsewhere.

\end{section}

\begin{section}{Conclusion }

In this paper, we extend our previous NRG study on the quadratic coupling spin-boson model to the dynamical correlation function of $\sigma_x$, {\it i.e.}, $C_{\sigma_x}(\omega)$. We investigate the evolution of $C_{\sigma_x}(\omega)$ along two paths on the $\epsilon$-$\alpha$ plane for a generic sub-Ohmic bath with $s=0.3$. Our results are presented and compared with two other correlation functions studied before, $C_{\sigma_z}(\omega)$ and $C_{X}(\omega)$. We find that $C_{\sigma_x}(\omega)$ has a qualitatively similar behavior as $C_{\sigma_z}(\omega)$, {\it i.e.}, $C_{\sigma_x}(\omega) \propto \omega^{1+2s}$ in the weak-coupling regime $\alpha < \alpha_c$ and $C_{\sigma_x}(\omega) \propto \omega^{1-2s}$ at $\alpha = \alpha_c$. The high frequency part of both  $C_{\sigma_x}(\omega)$ and  $C_{\sigma_z}(\omega)$ has a prominent peak on top of a broad background. Close to the spin flip point, in contrast to the stable Rabi peak of $C_{\sigma_z}(\omega)$, the high frequency peak of $C_{\sigma_{x}}(\omega)$ is broadened significantly and the line shape changes qualitatively. This shows that the superconducting qubit system at the optimal working point has a shorter dephasing time at the spin flip line.

\end{section}

\vspace*{0.5cm}
\begin{section}{Acknowledgments }

This work is supported by 973 Program of China (2012CB921704), NSFC grant (11374362), Fundamental Research Funds for the Central Universities, and the Research Funds of Renmin University of China 15XNLQ03. We thank K. Yang for providing the $z$-average code.

\end{section}

\end{document}